\documentclass[superscriptaddress,aps,pre,twocolumn,showpacs,preprintnumbers]{revtex4-1}
\usepackage{amsmath}
\usepackage{times,mathptmx,amssymb}
\usepackage{graphicx}

\newcommand{\derpar}[2]{\frac{\partial #1}{\partial #2}}
\newcommand{\VEC}[1]{\mathbf{#1}}
\newcommand{\HAT}[1]{\hat{\mathbf{#1}}}
\newcommand{\tf} {t_{\text{f}}} 
\newcommand{\tfz} {t_{\text{f}0}} 
\newcommand{\tfu} {t_{\text{f}1}} 
\newcommand{\tfd} {t_{\text{f}2}} 
\newcommand{\dif} {{\rm d}}

\begin{document}

\title{Effective two-dimensional model for granular matter with phase separation}

\author{Dino Risso}
\affiliation{Departamento de F\'{\i}sica, Universidad del  B\'{\i}o-B\'{\i}o, Concepci\'on,  Chile}
\author{Rodrigo Soto}
\author{Marcelo Guzm\'an}
\affiliation{Departamento de F\'{\i}sica, FCFM, Universidad de Chile, Santiago, Chile}
\pacs{45.70.Mg, 
05.20.Dd	
}

\begin{abstract}
Granular systems confined in  vertically vibrated shallow horizontal boxes (quasi two-dimensional geometry) present a liquid to solid phase transition when the frequency of the periodic forcing is increased. An effective model, where grains move and collide in two-dimensions is presented, which reproduces the aforementioned phase transition. The key element is that besides the two-dimensional degrees of freedom, each grain has an additional variable $\epsilon$ that accounts for the kinetic energy stored in the vertical motion in the real quasi two-dimensional motion. This energy grows monotonically during free flight, mimicking the energy gain by collisions with the vibrating walls and, at collisions, this energy is instantaneously transferred to the horizontal degrees of freedom. As a result, the average values of $\epsilon$ and the kinetic temperature are decreasing functions of the local density, giving rise to an effective pressure that can present van der Waals loops. A kinetic theory approach predicts the conditions that must satisfy the energy grow function to obtain the phase separation, which are verified with molecular dynamics simulations. Notably, the effective equation of state and the critical points computed considering the velocity--time-of-flight correlations differ only slightly from those obtained by simple kinetic theory calculations that neglect those correlations. 
\end{abstract}
\maketitle

\section{Introduction}

Granular media, placed in vertically vibrated quasi two-dimensional (Q2D) shallow boxes are excellent systems to study the non-equilibrium nature of these materials. Energy is injected into the system via collisions of the grains with the top and bottom walls. By this, grains gain energy into the vertical direction, which is later transferred to the horizontal degrees of freedom via grain--grain collisions.  This mechanism is also dissipative,  by the effects of friction and inelasticity. Hence, energy follows a well defined path, which violates the detailed balance, placing the system in strong non-equilibrium conditions. When the height of the box is smaller than two grain diameters, it is possible experimentally to follow the motion of all grains at the individual and collective scale, providing elements to describe these systems in detail and test different theoretical approaches of non-equilibrium matter. See Ref.~\cite{reviewMujicaSoto} for a review on the different properties of granular matter in this geometry.

One of the most remarkable phenomenon of this geometry is that when the vibration amplitude exceeds a certain threshold, a non-equilibrium transition takes place and the system separates between dense and dilute phases, where the later presents a solid-like crystalline structure~\cite{Olafsen1998,Prevost2004,Melby2005}. This transition has been characterized in detail, describing its coarsening~\cite{CoarseningAranson}, crystallization~\cite{Reis2006}, and caging dynamics~\cite{Reis2007}, the criticality of the transition and its universality~\cite{Castillo2012,Castillo2015,Guzman2018}, and the existence of a surface tension between the phases~\cite{Luu2013}, to mention a few properties. 

Similarly to the case of thick granular layers~\cite{Argentina2002, Meerson2002}, this phase separation is triggered by the negative compressibility of the effective 2D equation of state and it  was demonstrated experimentally that the pressure versus density curve reaches a plateau precisely at the solid--liquid phase coexistence~\cite{Clerc2008}. 
The origin of the negative compressibility is that in shallow boxes, the granular temperature is not a slow field but, rather, it is enslaved to the density. Denser regions dissipate more energy, resulting in that the temperature is a decreasing function of density, as was experimentally observed~\cite{Prevost2004}. Hence, the pressure that in principle depends on density and temperature, ends up depending only on  density. When the temperature decreases rapidly enough with density, the effective equation of state can present van der Waals loops, resulting in a phase separation~\cite{Argentina2002,Argentina2003,Cartes,Khain2011}.

The theoretical analysis of the Q2D geometry presents the difficulty that grain--grain collisions must be parametrized considering the geometrical constrains imposed by the condition that the box height $H$ is smaller than twice the particle diameter $\sigma$. To simplify the analysis, two dimensional models have been built where some mechanism is included at collisions to mimic the energy injection that results from the three dimensional model. In Ref.~\cite{Barrat}, the restitution coefficient is taken to be random with possible outcomes larger than one, thus injecting energy.  However, the model does not provide steady states because it lacks an energy scale and the total energy  performs a random walk.
In an another approach, keeping a fixed restitution coefficient, collisions are modified to add an extra velocity to the relative motion~\cite{DeltaModel,DeltaShear}. These granular models have the common feature that energy is injected and dissipated at collisions, both being governed by the collision rate. Hence, the stationary temperature is density independent, except for weak effects that result from velocity correlations \cite{DeltaModel}. As a result, the effective pressure is a monotonous function of density and no van der Waals loop can appear. In this manuscript, we propose a granular model where it emerges naturally that the injected energy depends on the local density, giving rise to van der Waals loops and the development of phase separation.

The manuscript is organized as follows. In Section~\ref{sec.model} the granular model is presented in detail, where the three dimensional dynamics is encoded in an additional scalar degree of freedom that accounts for the energy stored in the vertical direction. Section~\ref{sec.quali} analyzes the model using kinetic theory arguments, finding the conditions on the model which are necessary to develop van de Waals loops. The predictions are compared with molecular dynamics simulations, with excellent agreement. For a more detailed analysis, a formal kinetic theory approach is performed in Section~\ref{sec.kt}, where  correlations between velocities and the additional degree of freedom are included. The calculations show that these correlations can be safely discarded to obtain estimates of the equation of state with good accuracy. Finally, Section~\ref{sec.discussion} gives a discussion of the results.

\section{Effective two-dimensional model} \label{sec.model}
The most complex part in the description of Q2D granular systems is the  three-dimensional parametrization of collisions, where the condition $H\gtrsim\sigma$ imposes severe restrictions on the kinematics (see, for example \cite{BreyConfined}). In order to build descriptions which can be worked out in detail, we simplify the kinematics reducing the motion to two dimensions, where particles are disks that have equal mass $m$ and diameter $\sigma$.   To mimic the three-dimensional motion, in which particles gain vertical energy by collisions with the plates, we model each particle to have an additional degree of freedom $\epsilon$, which represents the vertical energy and   grows monotonically with time between collisions. At collisions, energy is transferred  to the horizontal degrees of freedom depending on the values of $\epsilon$ of the colliding particles and, immediately after, the values $\epsilon$ are reset to zero. In combination with the restitution coefficient $\alpha$, which is the responsible for energy dissipation, energy can have positive or negative variations at every collision and it is possible to reach steady states. The advantage of this model, compared to the $\Delta$-model in Ref.~\cite{DeltaModel,DeltaShear} is that the amount of energy that is injected at collisions depends on the particles flight time, hence on density. Indeed, at low densities, the mean flight time is large, allowing particles to reach large values of $\epsilon$, which is transferred at collisions, leading to large values of the horizontal granular temperature $T$. On the contrary, at large densities, the transferred energy is small and $T$ is therefore small. This monotonically decreasing relation between $T$ and density will lead, as shown below, to a non-monotonous equation of state with a van der Waals loop, giving rise to phase separation. That is, the density is what fixes the time scale and, therefore, the steady temperature.

During free flight, the internal energy grows with a law $\epsilon=G(t)$, depending on a monotonic increasing function $G$ that will be specified below. This energy is released to the in-plane degrees of freedom in a collision rule that preserves lineal momentum. The postcollisional velocities and internal energies are given by
\begin{equation}\label{eq.colrule}
\begin{split}
\VEC{c}_1' &= h_1(\VEC{c}_1,\VEC{c}_2,\HAT{n},\epsilon_1,\epsilon_2)=\VEC{c}_1 - Q \HAT{n}, \quad \epsilon_1' =0, \\
\VEC{c}_2' &=  h_2(\VEC{c}_1,\VEC{c}_2,\HAT{n},\epsilon_1,\epsilon_2)=\VEC{c}_2 + Q \HAT{n},\quad
\epsilon_2' =0,
\end{split}
\end{equation}
where $\HAT{n}$ is the unit vector pointing from particle 1 to 2, $Q>0$ is the transferred linear momentum, and $h_{1/2}$ are the collision rules written explicitly on the terms they depend on \cite{DeltaModel}. No rotation is considered and particles are smooth. 
There are several options that generalize the collision rule for inelastic hard spheres, taking into account the energy injection. Here, we opt for the following collision rule, which allows us to advance in the analytical calculations: $Q$ is chosen such that the energy balance  at a collision is
\begin{equation}
\Delta E\equiv \frac{m}{2}\left(c_{1}'^2+c_2'^2 - c_1^2-c_2^2 \right) = \epsilon_1+\epsilon_2 - \frac{1-\alpha^2}{4}mc_{12n}^2,\label{eq.changeE}
\end{equation}
where $c_{12n}=(\VEC{c}_1-\VEC{c}_2)\cdot\HAT{n}>0$ is the normal relative velocity prior to collision. This gives
\begin{equation}
Q = \frac{c_{12n}}{2}-\frac{1}{2}\sqrt{\alpha ^2 c_{12n}^2+\frac{4}{m}\left( \epsilon _1 +\epsilon _2\right)}. \label{eq.Q}
\end{equation}
When $\epsilon_1=\epsilon_2=0$, the collision rule reduces to that of the inelastic hard sphere model~\cite{PoschelKT}, while if $\alpha=1$, it is of an exothermic collision. 
Other options for the collision rule, respecting these limits give similar qualitative results to those presented in the next sections.

This granular model can be directly studied numerically using event driven simulations of hard disks in two dimensions~\cite{MRC,DeltaModel}, implementing the collision rule \eqref{eq.colrule}. 
 Two kind of numerical simulations are performed. First, small systems composed of $N=1000$ hard inelastic particles, with restitution coefficient $\alpha=0.9$, in a square box of dimensions $L_x=L_y=\sqrt{N/n}$, and densities in the range $n\sigma^2=0.05$ to $n\sigma^2=1.1$, are considered to study the stationary homogeneous states. This small system size disallows any possible phase separation due to the high energy cost in forming an interface and, therefore, permits to obtain the eventual van der Waals loop in the equation of state. Secondly, to study the phase separation, elongated systems of large aspect ratio are studied $\sigma\ll L_y\ll L_x=5L_y$ . In both cases, periodic boundary conditions are used in the two directions.

\section{Qualitative analysis} \label{sec.quali}
\subsection{General approach}
Before performing a detailed analysis using formal kinetic theory (see Section~\ref{sec.kt}), we proceed to analyze qualitatively the model, making some simple assumption on the distribution function, which will be justified a posteriori. The steady state is characterized by a vanishing rate of change of the two-dimensional kinetic energy~\eqref{eq.changeE}. The average must be done over collisions, meaning that is should be evaluated over precollisional states and with a collision rate proportional to $c_{12n}$. As usual in the framework of the Boltzmann or Enskog kinetic theories, it is possible to assume that the colliding particles have no velocity correlations and spatial correlations are taken into account via the pair distribution function at contact $\chi$ \cite{PoschelKT,SotoKT}. Here, 
we we will assume that $\chi$ conserves its equilibrium form, $\chi(n)=(1-7\pi n\sigma^2/64)/(1-\pi n\sigma^2/4)^2$ \cite{correlationSotoMareschal,Henderson}.
For the average, we need therefore the stationary distribution function $f_0(\VEC{c},\epsilon)$. As $\epsilon$ grows monotonically, instead of $\epsilon$, we can use the time-of-flight since the last collision $\tf$. Hence, the object under study is $f_0(\VEC{c},\tf)$. For small inelasticities, the equilibrium distribution can be used as a good approximation. However, even at equilibrium, natural correlations appear between the speed and $\tf$. Indeed, faster particles collide more frequently and, therefore, $\tf$ have smaller values than average. It is possible to obtain an explicit expression for $f_0$ at equilibrium, but it is too involved to be of practical use. The marginal distributions, for velocities and  times of flight, are however simple: the Maxwellian and exponential distributions, respectively. Hence, for the purpose of computing the collisional averages, we approximate 
\begin{align}
f_0(\VEC c,\tf ) = \frac{n}{\pi c_0^2 \tfz} e^{-c^2/c_0^2} e^{-\tf/\tfz}, \label{eq.feq}
\end{align}
where $\tfz=[n\chi(n) \sigma\sqrt{2 \pi}c_0]^{-1}$\ is the mean flight time. In Section~\ref{sec.kt} we will improve  the approximation~\eqref{eq.feq} including  $c$-$\tf$ correlations. 
Note that if we had worked in the $f(\VEC c,\epsilon)$ representation, an additional Jacobian factor would have been necessary in \eqref{eq.feq}.
Within this  approximation it is direct to obtain
\begin{align}
\langle \Delta E\rangle_\text{col}=2\langle\epsilon\rangle - (1-\alpha^2) T, \label{eq.Teq}
\end{align}
where $T=mc_0^2/2$ is the granular temperature, defined with $k_\text{B}=1$ as usual, and 
\begin{equation}
\langle \epsilon\rangle = \int_0^\infty d\tf\, G(\tf) e^{-\tf/\tfz}/\tfz. \label{eq.avgepsilon}
\end{equation}
Eq.\ \eqref{eq.Teq} gives implicitly the stationary temperature  in terms of the density. To solve it, we need to specify the internal energy growth model, $G(t)$. Once solved, the stationary temperature $T(n)$ can be inserted into the equation of state~\cite{PoschelKT,Luding2009}
$p=n T [1+ (1+\alpha) n\chi(n)\pi\sigma^2/4 ] $,
to finally get the effective pressure $\widehat{p}$, which is a function of density. For $\alpha\approx 1$, this expression simplifies to 
\begin{align}
\widehat{p}(n)=n T(n) [1+ n \chi(n)\pi\sigma^2/2].  \label{eq.eqstate}
\end{align}
The objective of this study is to show whether appropriate choices for the energy growth function $G$ can generate van der Waals loops in the effective equation of state.

\subsection{Energy growth models}
The first model to be considered (M1) is just a linear growth of the internal energy, $G(t) = \Lambda\, t$, for which we describe in detail the procedure used to obtain the effective equation of state. 
Eq.~\eqref{eq.avgepsilon} gives $\langle\epsilon\rangle=\Lambda\, \tfz$. This allows to solve for the stationary temperature
\begin{equation}
T_\text{st}=E_s\,\left[\frac{1}{1-\alpha^2}\frac{1}{\sqrt{\pi}\,n\sigma^2  \chi(n)}\right]^{2/3},
\end{equation}
which is in effect a decreasing function of density. Here, $E_s=m\sigma^2[\Lambda/(m\sigma^2)]^{2/3}$ is an energy scale. Substituting in Eq.~\eqref{eq.eqstate}, the effective pressure $\widehat p(n)$ for this model, becomes monotonously increasing with density and no phase transition  takes place. The reason for this is the mild growth rate: as  discussed in Ref.~\cite{Cartes},  temperature must decrease sufficiently rapid with density to generate a van der Waals loop. 

\begin{figure}[htb] 
\begin{center}
\includegraphics[width=0.9\columnwidth]{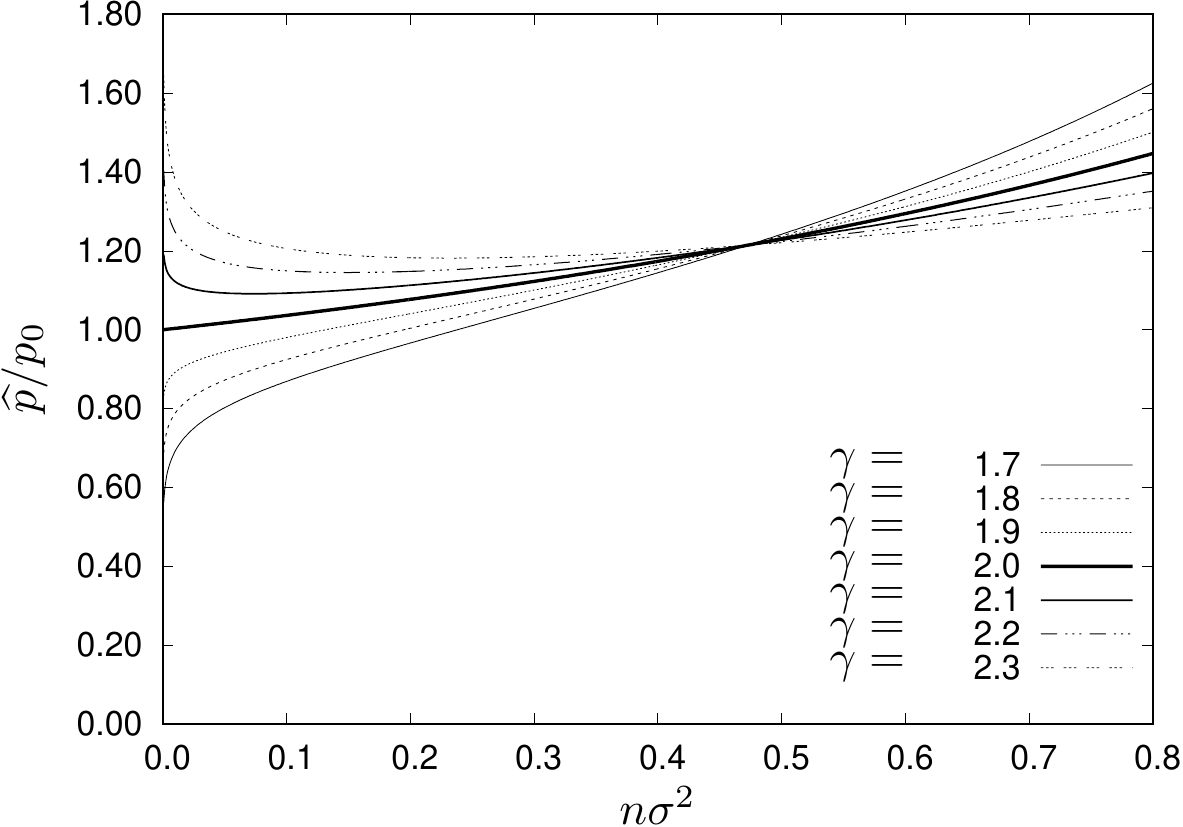}
\end{center}
\caption{Dimensionless pressure $\widehat{p}/p_0$ against density in model M2 for a set of values of $\gamma$  around  $\gamma=2$. The pressure scale is $p_0=E_s/\sigma^2 \{2\gamma!/[(1-\alpha^2)2\sqrt{\pi}]^\gamma\}^{2/(2+\gamma)}$.} 
\label{fig:presionDetail-M2}
\end{figure}

Hence, the next model to consider (M2) is a superlinear increase  $G(t) = \Lambda\, t^\gamma$, with $\gamma>1$.  Now results in $\langle\epsilon\rangle=\Lambda\, \gamma!\, \tfz^{\gamma}$, allowing to obtain the stationary temperature  
\begin{equation}
\label{eqn:Tvsgamma}
T_\text{st}=E_s\left\{\frac{1}{1-\alpha^2}\frac{2\gamma!}{[2\sqrt{\pi} n\sigma^2 \chi(n)]^\gamma}\right\}^{\frac{2}{2+\gamma}},
\end{equation}
where now the energy scale is $E_s=m\sigma^2[\Lambda/(m\sigma^2)]^{2/(2+\gamma)}$.
For small densities the temperature diverges as   $T_\text{st}\sim n^{-2\gamma/(2+\gamma)}$
and hence the effective pressure behaves like
\begin{equation}
\widehat p \sim n^{\frac{2-\gamma}{2+\gamma}}.
\end{equation}
For exponents $\gamma \leq  2$, the pressure is monotonically increasing (in the full range of densities). However, for  $\gamma > 2$,  it changes from  monotonically decreasing  
to monotonically increasing,  showing negative compressibilities for densities smaller than a threshold that depends on $\gamma$ (see Figure~\ref{fig:presionDetail-M2}). It is therefore a good candidate to model the phase transition. However, for $\gamma \geq 2$, the pressure does not vanish in the dilute limit, contrary to what is observed experimentally~\cite{Geminard2004,Clerc2008}. The origin of this anomalous behavior is that for low densities, the times of flight and, consequently, the average values of $\epsilon$ get large values, increasing therefore the stationary temperature, which diverges in the dilute limit. When computed, the product $p=nT$ does not vanish.

\begin{figure}[htb]
\begin{center}
\includegraphics[width=0.9\columnwidth,angle=0]{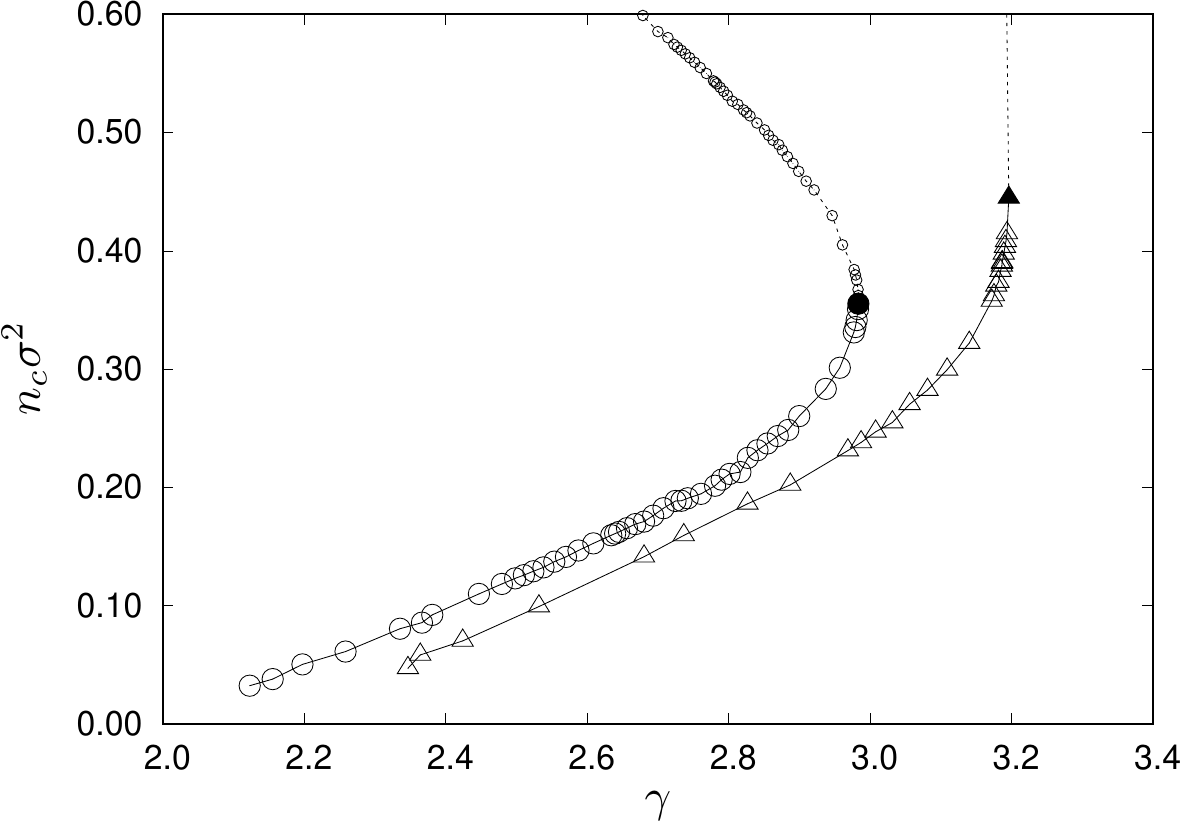}
\includegraphics[width=0.9\columnwidth]{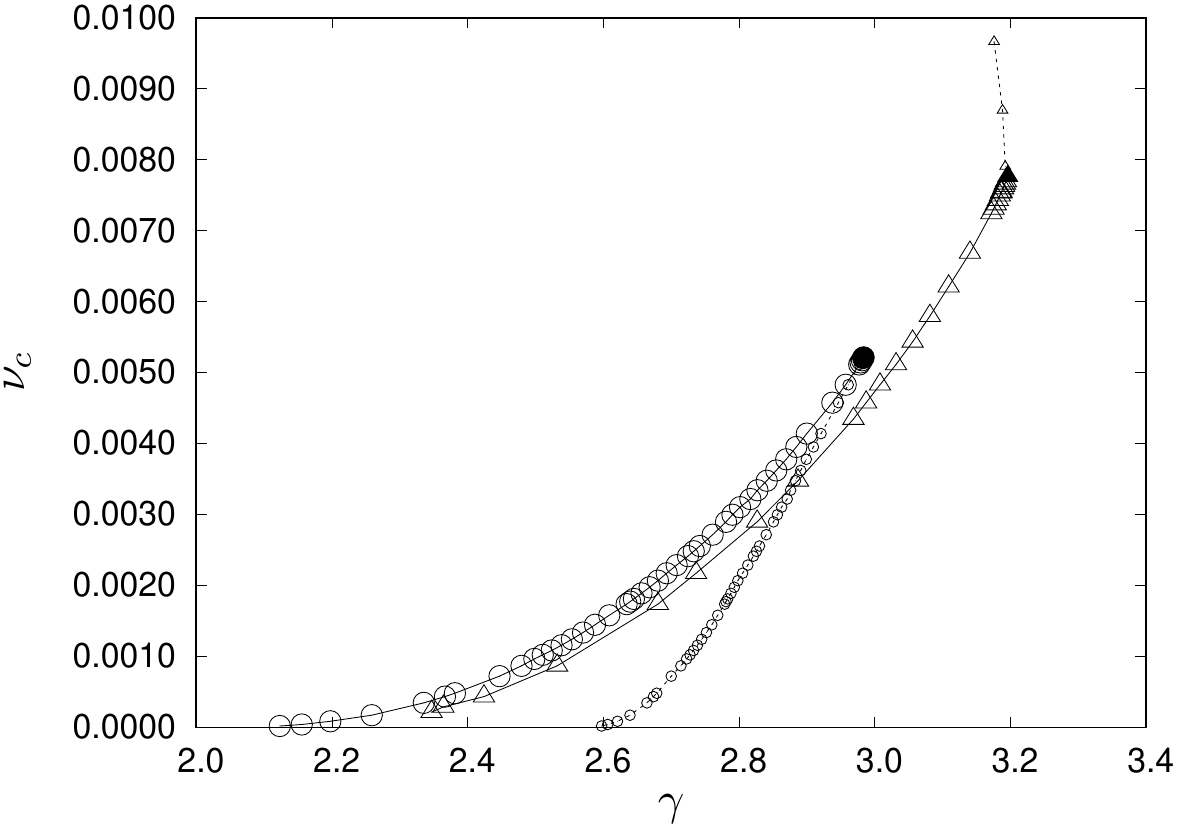}
\end{center}
\caption{
Critical  values  $n_c$ (top) and $\nu_c$ (bottom) for models M3 (circles) and M4 (triangles) with $\alpha=0.9$. 
The solid symbols indicate the maximum $\gamma$ where a critical point exists, with values $\gamma\approx2.98$, $n_c \approx 0.35/\sigma^2$, and $\nu_c\approx5.21\times 10^{-3}$  for model M3 and  $\gamma_c \approx3.20$, $n_c \approx 0.44/\sigma^2$, and $\nu_c\approx7.75\times 10^{-3}$ 
for model M4 with $\epsilon_0=10^{-4}$.
The light lines are visual guide lines. In small symbols with dashed lines, the high density branch of critical points.}
\label{fig.criticalParametersM3}
\end{figure}

To overcome this difficulty, we need to introduce a saturation on $\epsilon$, feature that is well justified experimentally when we consider the mechanisms by which the vertical energy grows in Q2D systems by collisions with the vibrating plates. Taking into account the superlinear growth, we model  $G(t) = \epsilon_{\infty} (1-e^{-t/\tau})^\gamma$, (M3). Based on the Q2D geometry in absence of gravity, the characteristic time  scales as  $\tau\sim (H-\sigma)/(A\omega)$ and the maximum energy as $\epsilon_{\infty}\sim m(A\omega)^2(1+\alpha^2)/(1-\alpha^2)$, where $A$ and $\omega$ are the amplitude and angular frequency of the vibration~\cite{TheisNico}. With this model  the stationary temperature saturates at a value $T_\text{max}=2\epsilon_{\infty}/(1-\alpha^2)$ for low densities  and the pressure effectively vanishes at zero density.
Computing $\langle\epsilon\rangle$ and defining $x=\tau/\tfz$, Eq.~\eqref{eq.Teq} can be written as
\begin{equation}
\frac{\Gamma(1+\gamma)\Gamma(x)}{\Gamma(1+\gamma+x)x} =\mu^2, \label{eq.M3}
\end{equation}
where $\Gamma$ is the gamma function. Eq.~\eqref{eq.M3} can be solved numerically for given values of 
$\mu =  \nu/[n\sigma^2\chi(n)]$, where $\nu$ is a dimensionless relaxation rate,
\begin{align}
\nu=\sqrt{\frac{1-\alpha^2}{8\pi}} \frac{\sqrt{m\sigma^2/\epsilon_\infty}}{\tau}.
\end{align}
In terms of these variables, the stationary temperature is $T_\text{st}=2\epsilon_\infty \mu^2 x^2/(1-\alpha^2)$, allowing to obtain the pressure using Eq.~\eqref{eq.eqstate}. 
For $\gamma\leq 2$ the pressure is monotonically increasing with density, independent of the value of $\nu$. Whereas, for $\gamma > 2$ a set of critical points exist $[n_c(\gamma),\nu_c(\gamma)]$  (see Fig.~\ref{fig.criticalParametersM3}), for which ${\rm d}{\widehat p}/{\rm d}n=0$ and ${\rm d}^2{\widehat p}/{\rm d}n^2=0$. When $\nu<\nu_c(\gamma)$, a van der Waals loop shows up around $n_c(\gamma)$.  Note that when $\gamma$ approaches 2, both the critical density and relaxation rate vanish. 
A second branch of critical points appears at  higher densities. It is necessary for topological reasons to separate the regions where pressure is monotonously increasing [$\nu<\nu_c(\gamma)$] to where it presents a van der Waals loop [$\nu>\nu_c(\gamma)$] and depends on the specific details of the model for densities close to packing.

\medskip

Therefore, the system is predicted to present a phase separation when prepared at the appropriate conditions. The simulations in an elongated box exhibit, as predicted, a phase separation and the dense phase orders in a crystalline cluster with triangular symmetry. However, the density of this phase does not saturate and finally, the simulation presents inelastic collapse \cite{McNamara}. This feature, which is not observed in experiments, where the dense phase is stable and presents fluctuations \cite{Prevost2004,Clerc2008}, is due to a singularity of the model M3.  Indeed, when the system approaches the close packing density $n_\text{CP}=2/(\sqrt{3}\sigma^2)$, the flight times go to zero and therefore the stationary temperature vanishes rapidly enough to compensate the density divergence in $\chi$. Combined with the  density divergent factor $\chi(n)$ in Eq.~\eqref{eq.eqstate}, the pressure goes to a finite value at close packing. That is, there is no pressure divergence at close packing and the van der Waals loop is not closed. Therefore, it is not possible to determine the steady state by performing an analog to the Maxwell construction with finite densities.

Experimentally, the inelastic collapse does not take place because particles and plates are not perfect and there is always a  remaining energy after collisions. Although small, this effect is enough to create a repulsive pressure that avoids the collapse. In the final model (M4), we can simulate this property by considering that there is always a small  energy $\epsilon_0\ll\epsilon_\infty$, such that the energy balance equation changes to 
\begin{equation}
\Delta E\equiv \frac{m}{2}\left(c_{1}'^2+c_2'^2 - c_1^2-c_2^2 \right) = \epsilon_1+\epsilon_2 + 2 \epsilon_0- \frac{1-\alpha^2}{4}mc_{12n}^2. \label{eq.changeEM4}
\end{equation}
Now the transferred momentum $Q$ is given by Eq.~\eqref{eq.Q} substituting $\epsilon_{1/2}$ by $\epsilon_{1/2}+\epsilon_0$.
Then, even for a very rapid sequence of collisions, there is always a finite energy that is transferred to the horizontal degrees of freedom leading to a finite stationary temperature. 
The pressure is determined as in model M3, adding the term $(\epsilon_0/\epsilon_\infty)x^{-2}$ to the left hand side of Eq.~\eqref{eq.M3}. 
For $\epsilon_0/\epsilon_\infty=10^{-4}$ a set of critical points are obtained for $2<\gamma<3.196$, shown in Fig.~\ref{fig.criticalParametersM3}. 
Other values of $\epsilon_0$ only change the critical values but the whole picture remains qualitatively valid. 
Similarly to model M3, a second branch of critical points appears.

\begin{figure}[htb]
\begin{center}
\includegraphics[width=0.9\columnwidth]{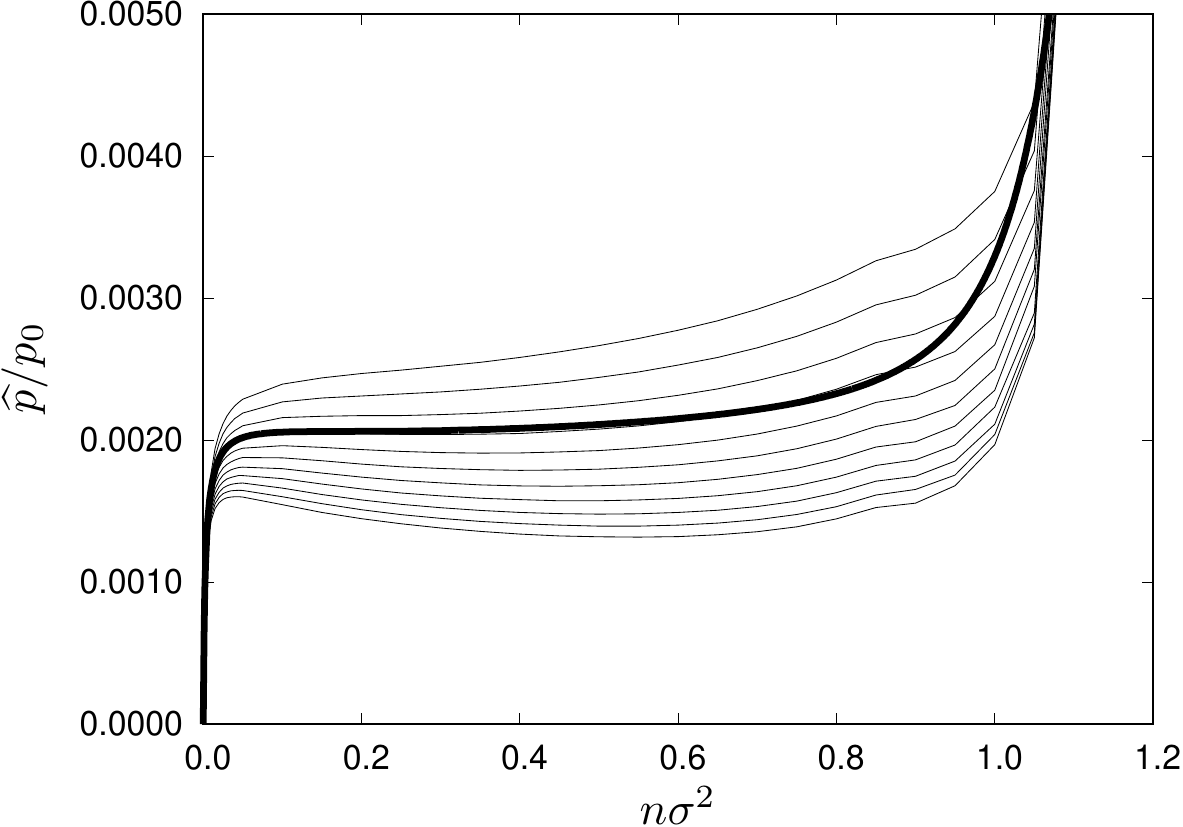}
\end{center}
\caption{
Simulational results for pressure versus density ---in the case of small box--- when using  model M4 with $\alpha=0.9$,  $\nu=0.00267$, and $\epsilon_0=10^{-4}$. The thin lines corresponds to a set of $\gamma$ values above and below the critical $\gamma_c\approx 2.8$. From top to bottom, $\gamma=2.5, 2.6, \dots ,3.5$. 
The pressure is scaled with $p_0= \frac{2\epsilon_\infty}{(1-\alpha^2)}$ and, in order to compare, in solid  line the theoretical prediction for the critical $\gamma$ value. 
\label{fig.pressuremodel4}
}
\end{figure}

\begin{figure}[htb]
\begin{center}
\includegraphics[width=0.95\columnwidth]{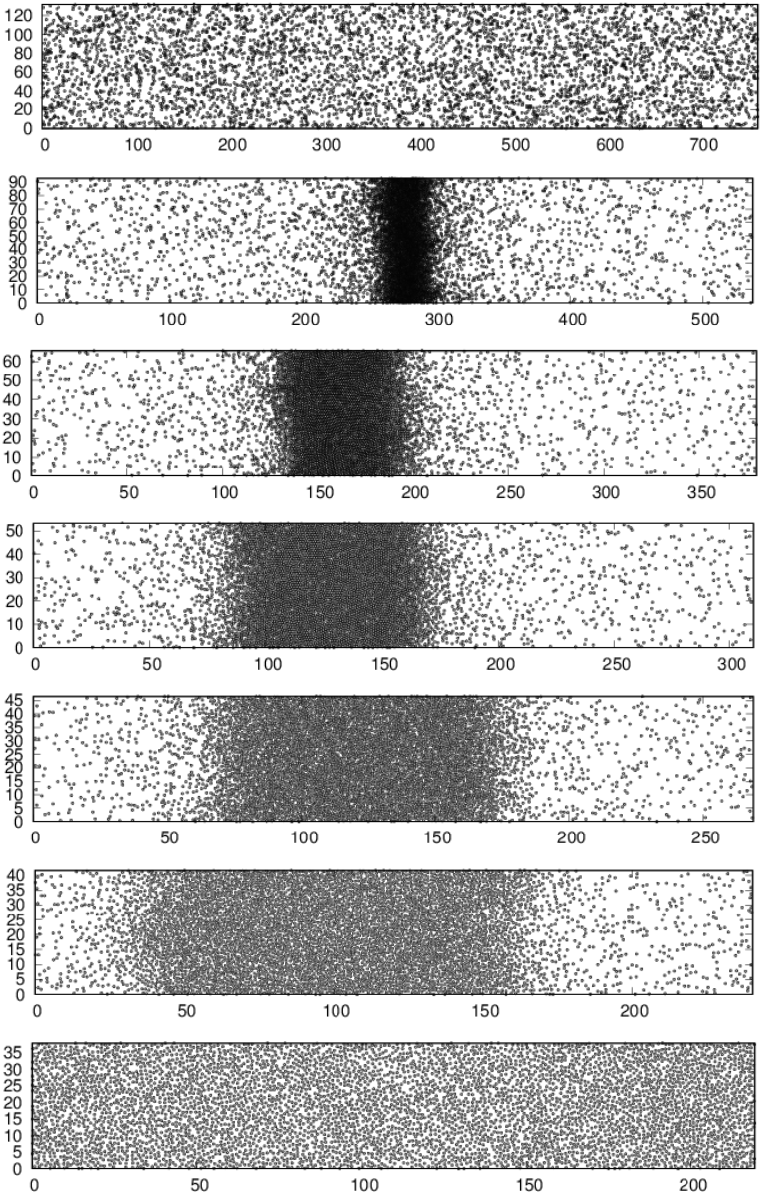}
\end{center}
\caption{Instantaneous configurations of simulations of the model M4 in the steady state with $N=5000$, $\alpha=0.9$,  $\nu=0.00267$, $\gamma=3.5$, and $\epsilon_0=10^{-4}$. From top to bottom the global densities are $n\sigma^2=0.05, 0.1, 0.2, 0.3, 0.4, 0.5$ and $0.6$. Note that for different densities the system sizes are different. In the density range $0.1\leq n\sigma^2\leq0.5$ the system separates in two phases.}
\label{fig.snapshots}
\end{figure}

Figure~\ref{fig.pressuremodel4} shows the pressure computed in simulations of a small box for different values of $\gamma$ and $\nu$. For comparison, the predicted pressure is also displayed, showing a very good agreement with simulations except for very high densities where the expression used for $\chi$ is not valid \cite{commentChi}.
To verify that indeed under the adequate conditions a phase separation takes place, molecular dynamics simulations are performed. 
 Figure~\ref{fig.snapshots} shows stationary configurations for $\alpha=0.9$, $\nu=0.00267$, and $\gamma=3.5$, which  is deep in the developed region where the Van deer Waals loop appears (see Figure~\ref{fig.criticalParametersM3}).

\section{Kinetic theory} \label{sec.kt}

\subsection{Kinetic equation}

In Section~\ref{sec.quali} it was shown that, discarding the velocity--time-of-flight correlation, a phase separation takes place when the internal energy grows according to model M4. However, discarding correlations is an uncontrolled approximation. In order to quantify the effect of these correlations, we proceed to describe model M4 with the tools of kinetic theory.
Assuming as usual that there are no pre-collisional velocity correlations, it is possible to write an Enskog--Boltzmann equation for the distribution function 
$f(\VEC r, \VEC c_1, \tfu)$, 
\begin{align}
\derpar{f}{t} + \VEC c_1\cdot\derpar{f}{\VEC r} +
\derpar{}{\tfu}{\left[\theta(\tfu) G'(\tfu)f\right]}  =J[f], 
\label{eq.enskog.tv}
\end{align}
where  $J=J_+-J_-$ is the collisional term, giving the rate of change of the distribution function by collisions, which is written adopting the form used in Ref.~\cite{DeltaModel},
\begin{widetext}
\begin{align}
J_+ &\equiv  \sigma \chi \int f\left(1^*\right)f\left(2^*\right) |\VEC{c}_{12}^*\cdot \HAT{n}|\theta \left(\VEC{c}_{12}^*\cdot\HAT{n} \right) \delta \left( \VEC{c}_1-h_1\left(1^*,2^*,-\HAT{n}\right) \right)\delta \left( \VEC{c}_2-h_2\left(1^*,2^*,-\HAT{n}\right) \right)\delta(\tfu)\delta (\tfd) \dif\HAT{n}\, \dif2\, \dif1^* \dif2^*,\\
J_- &\equiv \sigma \chi \int f(1)f(2)\ |\VEC{c}_{12}\cdot \HAT{n}|\theta \left(\VEC{c}_{12}\cdot \HAT{n} \right)\dif\HAT{n}\, \dif2 ,
\end{align}
\end{widetext}
where $i\equiv(\VEC c_i,t_{\text{fi}})$, $\dif i\equiv \dif^2c_i\,\dif t_{\text{fi}}$, and $\chi$  accounts for the configurational correlations at finite densities. The first two terms of the left hand side of Eq.~\eqref{eq.enskog.tv} are the usual streaming terms, while the third term accounts for the steady growth of $\tf$, where the Heaviside step function $\theta$ indicates that $\tf$ is always positive. $J_+$  is the Boltzmann gain term, where the Dirac deltas indicate that after collisions, particles have $\tf$ reset to zero and  the velocities are those that result when applying the collision rule~\eqref{eq.colrule}. $J_-$ is the usual loss term where the integrations should be done over velocities and times of flight of the partner particle.  For the sake of simplicity we have written the collision terms of a homogeneous system, but the extension to an inhomogeneous case is straightforward. 
For this equation to properly conserve mass, one must have that $\theta'(x)=\delta(x)$ consistently. We have chosen the convention that $\theta(0)=1$ and, hence, $\int_0^\infty \delta(x) \dif x=1$. Despite the apparent complexity of $J$, the collision integrals, used below to compute the rate of change of average quantities, are simple and direct. For any function $\psi(\VEC c, \tf)$, \cite{DeltaShear,GarozBritoSoto}
\begin{multline}
\int \psi(\VEC c_1, \tfu) J[f]\, \dif1 = 
\frac{1}{2}\sigma \chi \int f(1)f(2)\ |\VEC{c}_{12}\cdot \HAT{n}|\theta \left(\VEC{c}_{12}\cdot \HAT{n} \right)\\
 \times\left[\psi(1')+\psi(2')-\psi(1)-\psi(2)\right]
 \dif\HAT{n}\, \dif1\, \dif2,\label{eq.colintegral}
\end{multline}
with $1'$ and $2'$ the postcollisional states of the particles given by Eq.\ \eqref{eq.colrule}.

\subsection{Stationary distribution and equation of state in the dilute regime}
Similarly to the case of undriven granular matter, it is not possible to obtain an exact solution of the Boltzmann--Enskog equation, even in the steady state for $\alpha<1$~\cite{PoschelKT}. 
We recall that the distribution function will be used to compute the temperature and pressure. Hence, we look for expressions in the thermal sector, i.e.~when $c\sim\sqrt{T/m}$, in which case it is possible to take into account the $c$-$\tf$ correlations with a polynomial  approximation
\begin{align}
f_0(\VEC c,\tf ) = \frac{n}{\pi c_0^2 \tfz} e^{-c^2/c_0^2} e^{-\tf/\tfz}\left[1+ 4a(c^2 -c_0^2)(\tf - \tfz)\right] ,\label{eq.distrfnGrad}
\end{align}
where $a$ quantifies the strength of the correlations and the factor 4 has been introduced for later convenience. 

In equilibrium ($\alpha=1$ and $G=0$), the temperature, hence $c_0$, is set by the initial condition and $\tfz$ is the corresponding mean flight time. The correlation intensity $a$ is computed as in the Grad moment method, by imposing that in the kinetic equation  the average of $c^2 \tf$ is stationary. Multiplying Eq.\ \eqref{eq.enskog.tv} by $c^2 \tf$, integrating over velocities and time-of-flight, and using Eq.\ \eqref{eq.colintegral} gives $a_\text{eq}=-1/35$.  This negative value indicates, as anticipated, that particles with larger speed tend to have smaller times-of-flight because of their larger collision frequency. Note, however, that the value is small and it will be shown below that weak modifications to the pressure are obtained by including these correlations.

We now proceed to consider the inelastic case ($\alpha<1$) in the dilute regime, considering first  the internal energy growth law of model M3. The parameters of the distribution function~\eqref{eq.distrfnGrad} are obtained imposing now that the averages of $c^2$, $\tf$, and $c^2\tf$  are stationary, resulting respectively in the following equations
\begin{align}
\frac{\Gamma (x)\Gamma(1+\gamma)}{\Gamma(1+\gamma+x)}\left\{ \frac{1-a}{x}  - a \left[ \Upsilon(x) - \Upsilon(1+\gamma+x) \right] \right\}&= \mu^2,
\label{Balancec2} \\
c_0\sigma n  \tfz (1+a)\sqrt{2\pi}&=1,
\label{Balancetv} \\
c_0\sigma n  \tfz (5+39 a)\sqrt{2\pi}&=4,
\label{Balancec2tv}
\end{align}
where  $\Upsilon$ is the digamma function.

The last two equations, which remarkably do not depend on the specific choice of the energy growth model, give $a=a_\text{eq}=-1/35$ and $c_0=1/[\sigma n  \tfz(1+a)\sqrt{2\pi}]$. Using the value of $a$, the first equation gives, implicitly, $x$ as function only of $\mu$ and the exponent $\gamma$. 
The pressure, under the assumption of low densities is simply that of an ideal gas $\widehat p=nmc_0^2/2=p_0 x^2\mu/(1+a)^2$, where $p_0=\sqrt{\epsilon_\infty/[2\pi(1-\alpha^2)m\sigma^2]}m/\tau$. 
The rescaled pressure $\widehat p/p_0$ is only a function of $\mu$ and $\gamma$.  
For values of $\gamma$ lower than 2.3 a monotonic increasing behavior is obtained, while for  $\gamma>2.3$, there are values of $\mu$ in which the compressibility is negative, revealing a phase separation. 
It is worth pointing out that, as expected for model M3, the pressure vanishes as the density goes to zero and, also, it does not diverge at close packing, which it is also not possible to capture in the framework of the Boltzmann equation.
Notably, the $c$-$\tf$ cross correlations, which can be suppressed by imposing $a=0$, have weak effects on the behavior of the pressure.  
This result justifies, a posteriori, the factorization approximation performed in Section~\ref{sec.quali}.

\subsection{Dense regime}
\begin{figure}[htb]
\centering
\includegraphics[width=0.9\columnwidth]{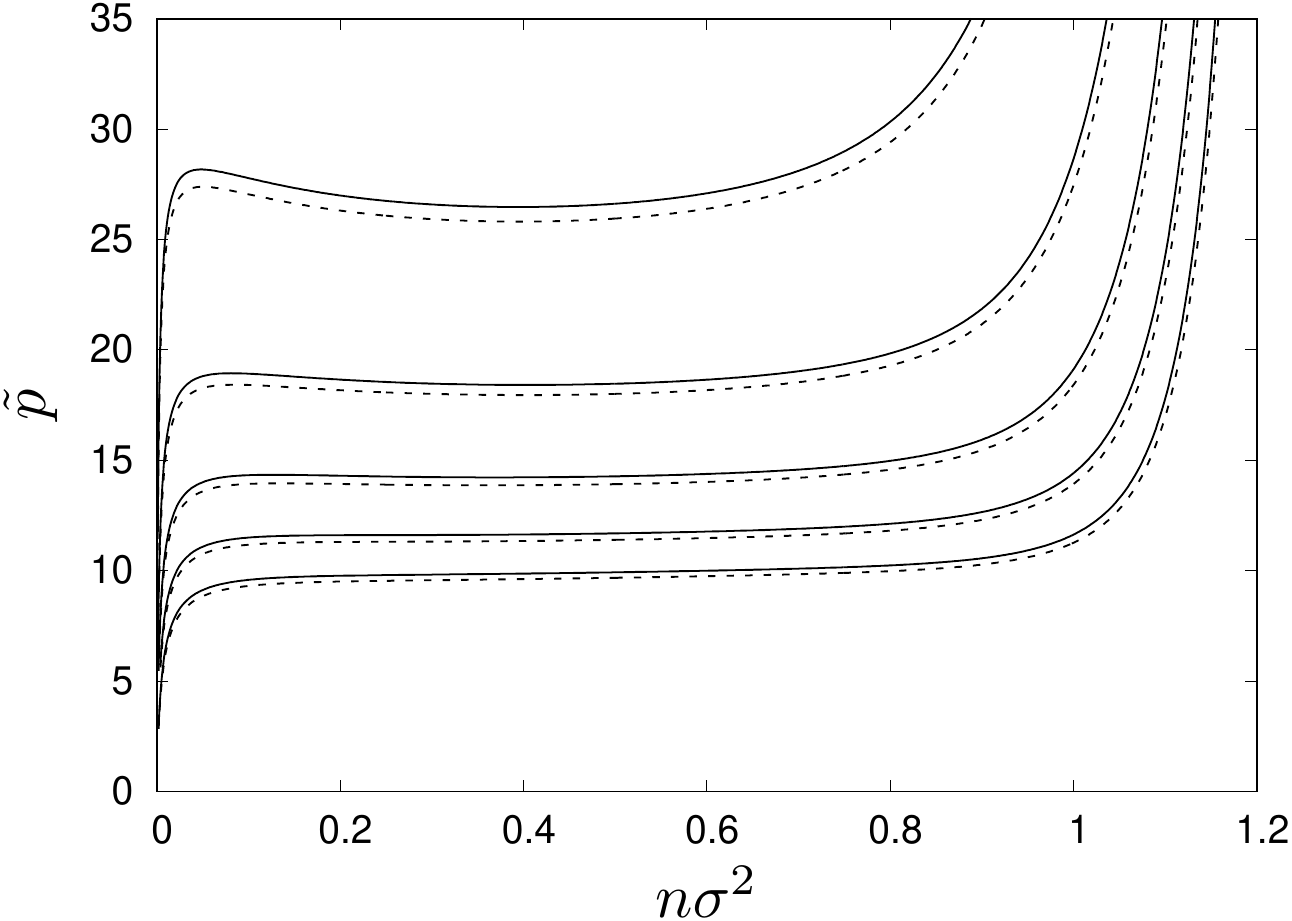}
\caption{Dimensionless pressure $\widetilde p$ of model M4 as a function of the dimensionless density $n\sigma^2$ for $\gamma=3$, $\epsilon_0/\epsilon_{\infty}=10^{-4}$, and  different values of $\nu$, from top to bottom 0.02, 0.03, 0.04, 0.05, and 0.06. Solid lines give the full result of kinetic theory and dashed lines are obtained neglecting the speed--time-of-flight correlations ($a$ set to 0).}
\label{HighDensityPressure1}
\end{figure}

For dense systems, we have already discussed in Section~\ref{sec.quali} that model M3 does not show the correct behavior near close packing and M4 should be used instead.  At the level of the kinetic equation, the factor $\chi$ should be considered on the balance equations for $c^2$, $\tf$ and $c^2\tf$. Eqs.~\eqref{Balancetv} and \eqref{Balancec2tv} only change by the addition of a global prefactor $\chi$ on the left hand side, resulting in the same value for $a$, while $c_0$ is simply divided by $\chi$. The balance equation 
for $c^2$, Eq.\ \eqref{Balancec2}, which does depend on the energy growth model, is modified by adding the term $(\epsilon_0/\epsilon_\infty)x^{-2}$ on the left hand side. 
The structure of the equation of state is  more complex  than in the dilute regime, for  the dimensionless pressure $\tilde p\equiv\widehat p \tau\sigma/\sqrt{\epsilon_\infty m}$ depends now on four dimensionless parameters: $\mu$, $\epsilon_0/\epsilon_\infty$,  $\nu$, and $\gamma$. The effect of $\epsilon_0/\epsilon_\infty\neq0$ is simply to produce a divergence of the pressure at close packing and we can fix it to an arbitrarily small value. 
As in Section~\ref{sec.quali}, van der  Waals loops appear for  $0<\nu<\nu_c(\gamma)$, while for higher values of $\nu$ the effective pressure is monotonous. 
Figure \ref{HighDensityPressure1} shows $\tilde p$ for $\gamma=3$ (which shows negative compressibility in the low density regime) for increasing values of $\nu$, comparing the results with and without speed--time-of-flight correlations. Notably, the effect on the pressure of the cross correlations is weak  as well as on the critical values $\nu_c(\gamma)$ and $n_c(\gamma)$ (not shown).

\section{Discussion} \label{sec.discussion}
It has been shown that the phase transition observed in vertically vibrated quasi two-dimensional granular fluids can be described by a simple effective two-dimensional model. The major advantage is that it is not necessary to treat in detail the three dimensional collision in the constrained geometry as the model is purely two-dimensional. The vertical motion is included by adding an additional scalar degree of freedom that gives the kinetic energy stored in this direction, which is  released to the horizontal degrees of freedom upon collisions. This dimensional reduction allows to perform detailed analytical calculations and build kinetic theories which can be used to compute the pressure and other quantities.
There are several conditions that the vertical kinetic energy growth function must obey in order to obtain an equation of state that presents a van der Waals loop and, hence, a phase transition can appear. First, the energy growth must be superlinear in order that the stationary temperature decreases rapidly when increasing the density, resulting in an effective pressure that can show negative compressibilities. Second, the vertical energy must saturate at long times to get finite temperatures and vanishing pressure for vanishing density. Finally, at collisions a small residual energy must be kept, mimicking the microscopic asperities of the granular system, which results in a finite granular temperature and a divergent pressure at close packing. All these conditions, except for the superlinear growth, are well justified from experimental conditions. It would be interesting to investigate in the future if the average energy growth in quasi two-dimensional systems follows this or similar laws as well as relate the model parameters with the grains' inelasticities\cite{Reyes2008} and with the amplitude and frequency of the vibration. 

In the full quasi two-dimensional case, depending on various forcing parameters, the dense phase can present different crystalline phases by forming one or two interlaced layers~\cite{Melby2005}. For the present model, the dense phase always shows a triangular crystalline order. More phases are not possible to obtain for hard two-dimensional particles, but it could be possible to include additional degrees of freedom to model these phases and the defects that can show up. 

\acknowledgments
This research was supported by the Fondecyt Grant No.~1140778. M.G.~acknowledges the Conicyt PFCHA Magister Nacional Scholarship 2016-22162176. D.R.~acknowledges the DIUBB Grant 18227 4/R.

\end{document}